\documentstyle[amssymb,secnumtab,epsfig]{fbssuppl}

\title{Few nucleon dynamics in a nuclear medium}
\author{Michael Beyer\thanks{{\it E-mail address:} 
beyer@darss.mpg.uni-rostock.de}}
\institute{Fachbereich Physik, Universit\"at Rostock, D-18051 Rostock, Germany}

\newcommand{\CG}{{\cal G}}
\newcommand{\gS}{\Sigma}
\newcommand{\ga}{\alpha}
\newcommand{\gb}{\beta}
\newcommand{\gc}{\gamma}
\newcommand{\gd}{\delta}

\begin{document}

\maketitle
\begin{abstract}
  Few body methods are used in many particle physics to describe
  correlations, bound states, and reactions in strongly correlated
  quantum systems. Although this has already been recognized earlier,
  rigorous attempts to treat three-body collisions have only been done
  recently. In this talk I shall give examples and areas where
  few-body methods have been and might be of use in the future.
\end{abstract}

\section{Introduction}
Describing an ensemble of many particles (fermions/bosons) becomes
challenging and interesting as soon as interactions (e.g. Coulomb or
strong interaction) are considered.  Examples for Coulombic systems
are {\em ionic plasmas} as they occur in the sun and stars, {\em
  liquid metals} and {\em electron-hole plasmas}. {\em Nuclear matter}
and the {\em quark gluon plasma} are examples of strongly interacting
systems. Because of the interaction it is not possible to treat even
single particle dynamics without regarding effects of the other
particles. Furtheron the system may be in equilibrium or out of
equilibrium, depending on the boundary conditions imposed.

\begin{figure}[tb]
\begin{minipage}{0.48\textwidth}
\psfig{figure=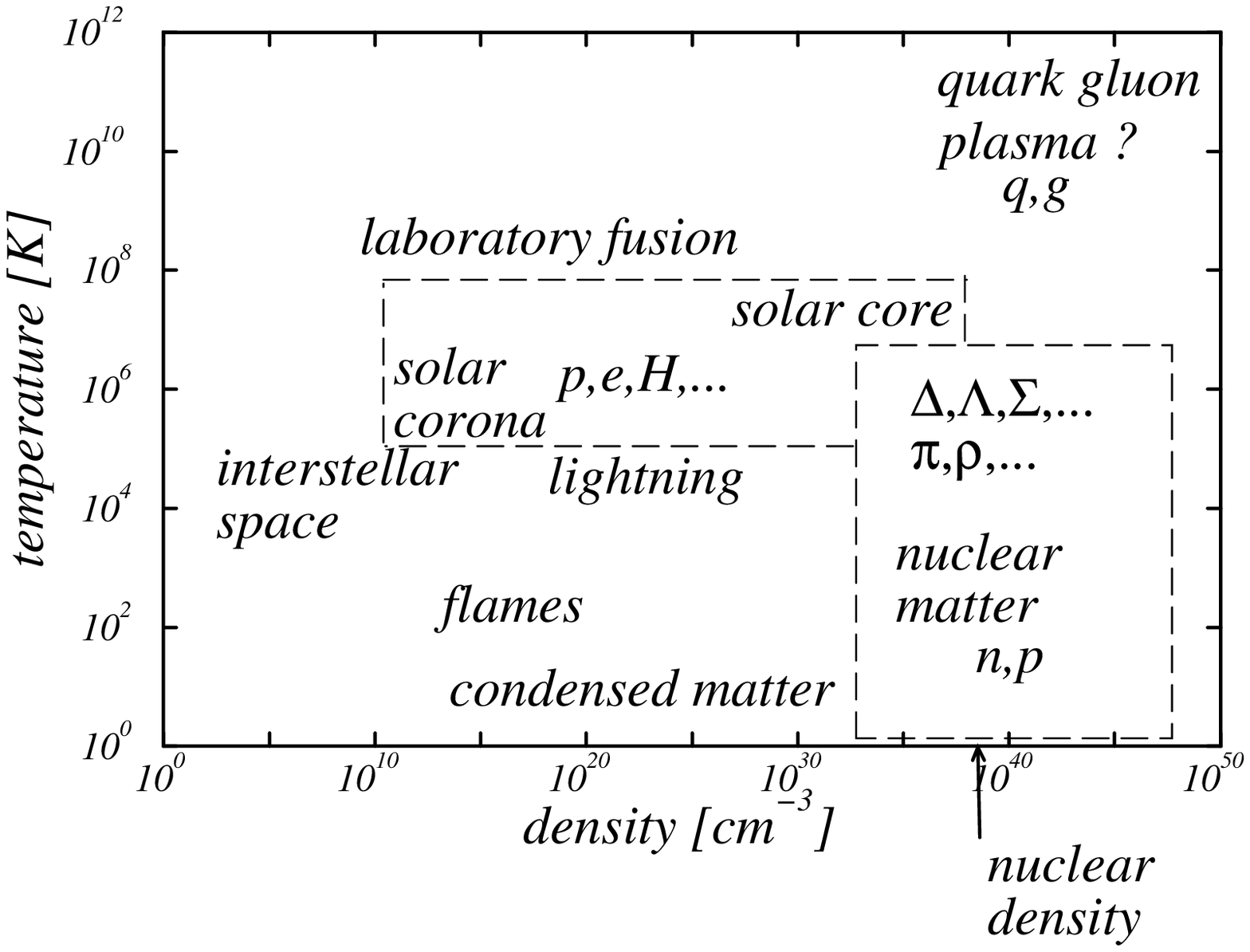,width=\textwidth}
\caption{\label{fig:ph_proc} Density temperature phase diagram of matter.}
\end{minipage}
\hfill
\begin{minipage}{0.48\textwidth}
\psfig{figure=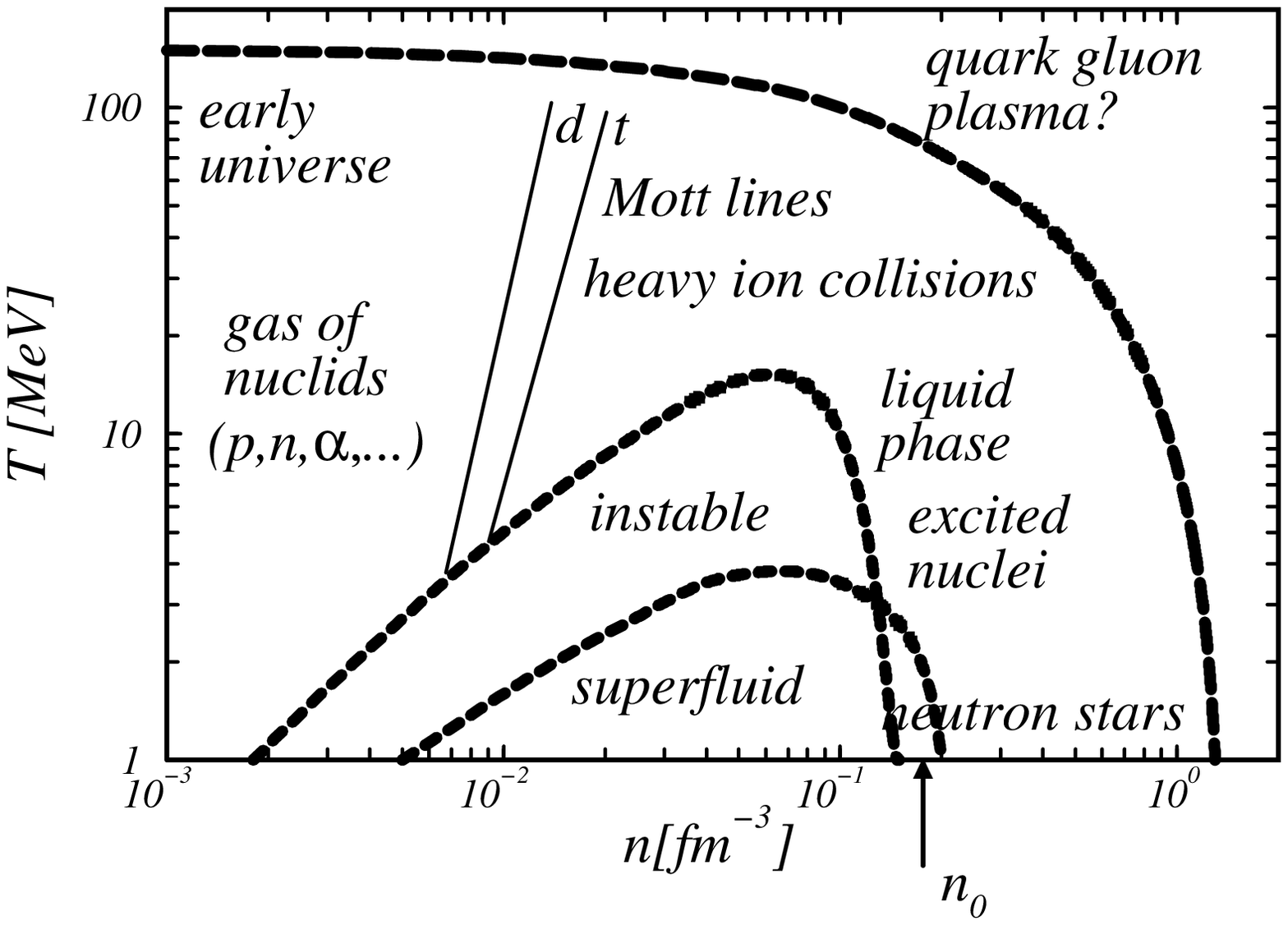,width=\textwidth}
\caption{\label{fig:nt_proc} Density temperature plane of nuclear matter.}
\end{minipage}
\end{figure}

The density temperature planes of matter and nuclear matter are shown
in Figs. \ref{fig:ph_proc} and \ref{fig:nt_proc}. The phase diagram of
nuclear matter turns out to be very rich. In particular the superfluid
phase reflecting strong pairing is relevant for the structure of
neutron stars~\cite{pin92}. At lower densities bound states occur.
This part of the phase diagram may be accessed in the laboratory
through heavy ion collisions at intermediate energies. The conditions
for the formation of bound states are reached in particular during the
final stage, where the nuclear density drops below the Mott density.

\section{Theory}
Quantum statistics provides powerful methods to tackle the many
particle systems. Here I follow the Green function
formalism~\cite{kad62}, which is convenient to introduce
few-body methods. Let the Hamiltonian of the system be given by
\begin{equation}
H(t) = \sum_1 H_0(1) \psi^\dagger_1(t) \psi_1(t)
+\sum_{12} V_2(12,1'2')
\psi_1^\dagger(t)\psi_2^\dagger(t)\psi_{2'}(t)\psi_{1'}(t),
\end{equation}
where $\psi_1(t)$ denotes the Heisenberg operator of the particle with 
quantum numbers $s_1$, $k_1$, etc. for spin, momentum
etc. The one particle Green function is defined by
\begin{equation}
i\CG_1(1,1')=\langle T \psi_1(t)\psi^\dagger_{1'}(t')\rangle
\equiv {\rm Tr}\{\rho_0T \psi_1(t)\psi^\dagger_{1'}(t')\},
\label{eqn:green}
\end{equation}
where averaging is due to the density operator $\rho_0$. For an open
system in thermodynamical equilibrium the extremum condition for the
entropy leads to the following expression for the quantum grand canonical
density operator,
\begin{equation}
\rho_0=\frac{e^{-\gb(H-\mu\,N)}}{{\rm Tr}\{e^{-\gb(H-\mu\,N)}\}}.
\end{equation}
The temperature ($1/\gb=T$) and the chemical potential $\mu$ are 
the corresponding Lagrange parameters. Using the Heisenberg equation
for $\psi_1$ results in the following equation for $\CG_1$~\cite{kad62}
\begin{equation}
\CG_1(1,1') =
 \CG_1^{(0)}(1,1')
-\sum_{\tilde 1 2\bar 2}
\; \CG_1^{(0)}(1,\tilde 1) iV_2(\tilde 1 2,\bar 1\bar 2)
\; \CG_2(\bar 1\bar 2,1'2^+)\big|_{t_1=t_2}.
\label{eqn:hierarchie}
\end{equation}
The argument $2^+$ means that $t_{2^+} = t_2+0^+$ and 
\begin{equation}
\left(i\partial_{t_1}-H_0(1)\right)\CG_1^{(0)}(1,1') = \gd_{11'}.
\end{equation}
Eq.~\ref{eqn:hierarchie} shows already the basic problem of many
particle physics, the hierarchy. To find a useful truncation of the
$n+1$ particle Green function from the $n$ particle one, some notion
of the system is needed. Usually the hierarchy is truncated at the two
particle level, assuming binary collisions only. Three-particle
collisions have been treated at most approximately using Born (for
Coulombic systems) or impulse approximation (for nuclear matter).
This may not be sufficient, in particular, if explicit three-particle
processes are considered (e.g. such as cluster formation, where a
third particle is needed to achieve momentum conservation, etc.).

Eq.~\ref{eqn:hierarchie} may be formally decoupled by introducing the
self energy $\gS(1, \bar 1)$. 
\begin{equation}
\sum_{ \bar 1} \; \gS(1, \bar 1)\;\CG_1(\bar 1,1')\; =
- \sum_{\bar 1\bar 2 2}\;  iV_2(12,\bar 1\bar 2)
\; \CG_2(\bar 1 \bar 2,1'2^+).
\label{eqn:defS}
\end{equation}
In the simplest case the self energy may then be treated in mean field
(e.g. Hartree-Fock) approximation, i.e. $\CG_2\rightarrow
\CG_2^{(0)}=\CG_1(1,1')\;\CG_1(2,2') -\CG_1(1,2')\;\CG_1(2,1')$
viz. no two particle correlations (leading to an ideal gas of quasi
particles). Using $V_2(12,\bar 1\bar 2)=-V_2(12,\bar 2\bar 1)$ the
self energy $\gS^{HF}(1, 1')$ is given by 
\begin{equation}
\gS^{HF}(1, 1')
= i\sum_{2\bar 2} V_2(12,1'\bar 2)\CG_1(\bar 2,2^+),
\label{eqn:selfHF}
\end{equation}
which reduces to the standard expression, if a static potential is
used, $V_2(12,1'2')=\gd_{11'}\gd_{22'} V_2(12)$, since
$\CG_1(2,2^+)=\langle \psi^\dagger_2 \psi_{2}\rangle = f_2$ is the
one particle distribution function for fermions. For a given potential
Eq.~\ref{eqn:selfHF} constitutes a self consistent problem to
determine $\mu$ and $\gb$, solved by iteration
\begin{equation}
f_1\equiv f(\varepsilon) = \frac{1}{e^{\gb(\varepsilon - \mu)}+1}, 
\qquad \varepsilon = \frac{k^2}{2m}+\gS^{HF}(k).
\end{equation}
\begin{figure}[h]
\begin{minipage}{0.60\textwidth}
\psfig{figure=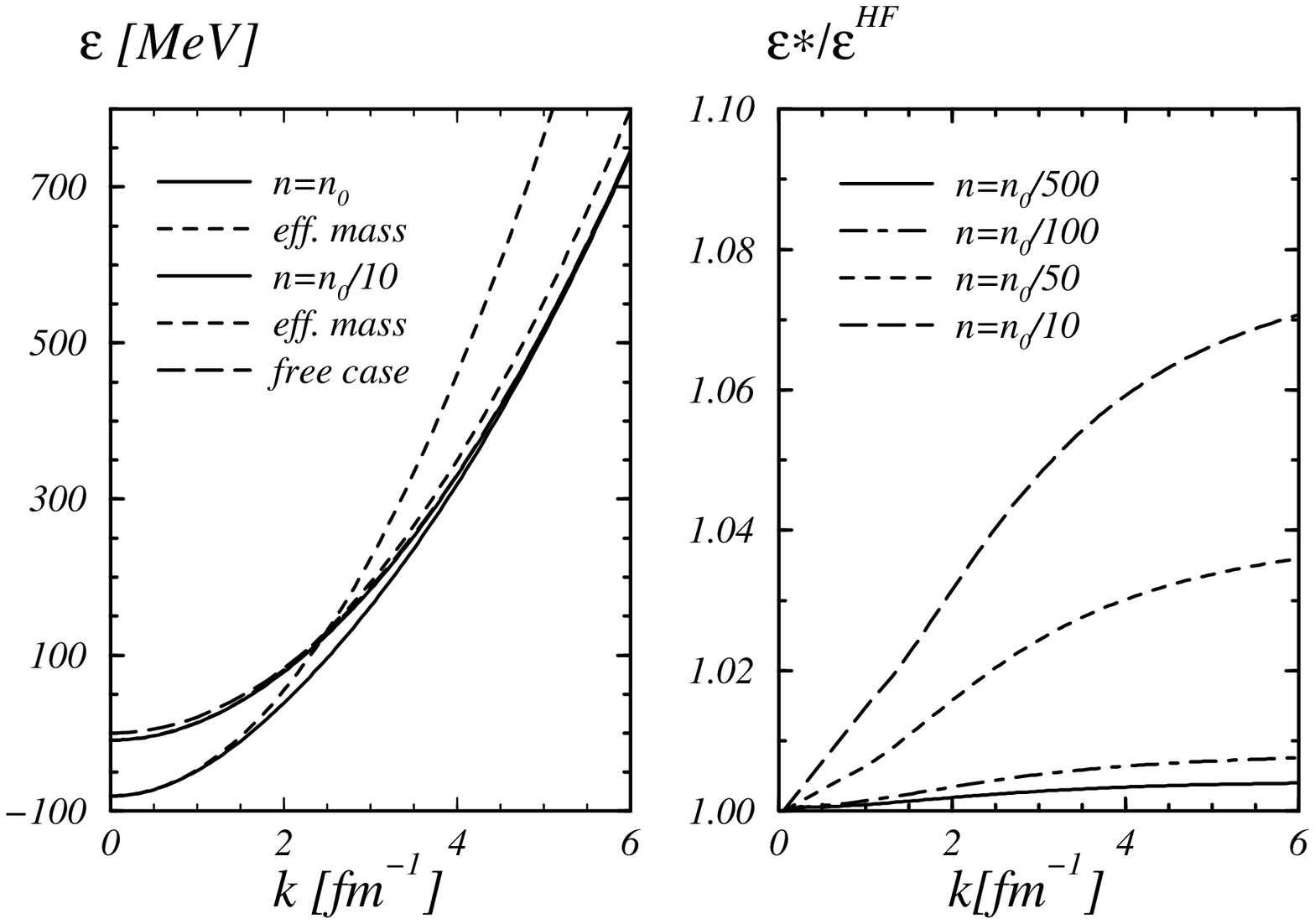,width=\textwidth}
\caption{\label{fig:self} The nucleon self energy (left
  side). Deviation of the effective mass approximation from the exact
  model result for small densities (right side).}
\end{minipage}
\hfill
\begin{minipage}{0.38\textwidth}
\psfig{figure=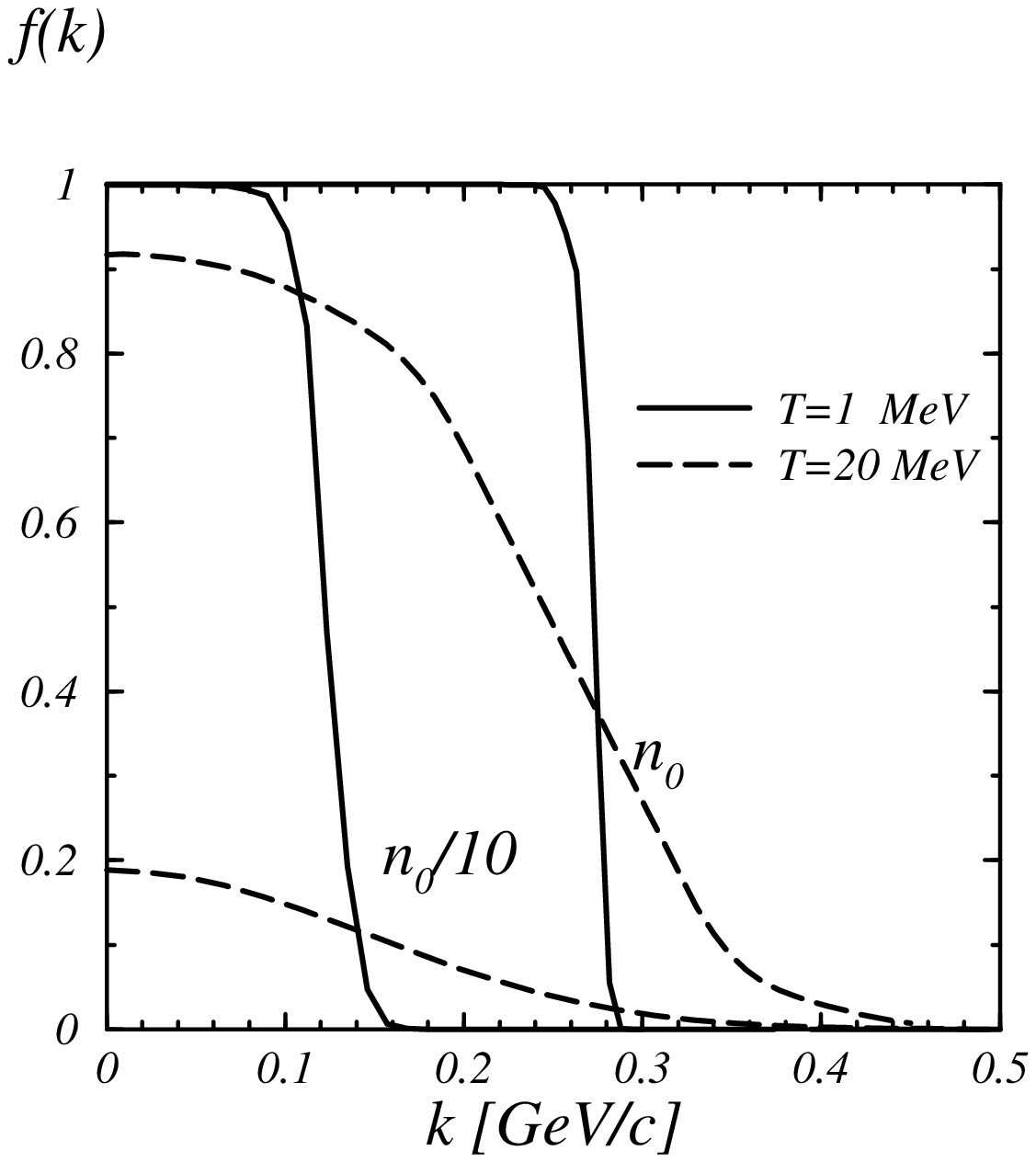,width=\textwidth}
\caption{\label{fig:fermi} One particle distribution function as a
  function of momentum $k$,
  $n_0=0.17$ fm$^{-3}$.}
\end{minipage}
\end{figure}

As a further simplification useful in later applications effective
masses may be introduced.  The self energy for the nucleon in a
nuclear medium of $T=10$ MeV is shown in Fig.~\ref{fig:self} along
with its effective mass approximation
$\varepsilon^*=k^2/(2m^*)+\gS^{HF}(k=0)$. The distribution function
$f(k)$ is shown in Fig.~\ref{fig:fermi}. Calculations are done using a
separable Yamguchi type potential (that will later be used for the
three body calculations). Finally, analytic continuation of the Green
function defined in Eq.~\ref{eqn:green} leads to the
Kubo-Martin-Schwinger boundary condition as $e^{itH}=e^{\gb H}$, i.e.
\begin{equation}
\CG_1(1,1')|_{t_1=0} = 
- e^{\gb\mu} \;\CG_1(1,1')|_{t_1=-i\gb}.
\label{eqn:KMS}
\end{equation}
As a consequence the time like component of the Fourier transform is
restricted to certain values only (Matsubara frequencies),
$\CG_1^{t_1-t_1'} \rightarrow G_1(z_\nu)$, where $z_\nu =
i\pi\nu/\gb+\mu$ and $\nu = \pm 1,\pm 3, \dots$ for
fermions. 

\section{Correlations}
A better treatment that goes beyond the
quasi particle approximation is provided, e.g. through the cluster
approximation that include correlations~\cite{roe83}. As a consequence 
the one particle spectral function $A_1(\omega)$, defined through
\begin{equation}
G_1(z_\nu)=\int \frac{d\omega}{2\pi}\frac{A_1(\omega)}{z_\nu-\omega}
\end{equation}
which is given by $A_1(\omega)=2\pi\,\gd(\omega-\epsilon)$ for the
quasi particle approximation, is more complicated, viz.
\begin{equation}
A_1(\omega) = \frac{2\,{\rm Im}\gS(\omega)}
{[\omega-E -{\rm Re}\gS(\omega)]^2 + [{\rm Im}\gS(\omega)]^2}.
\end{equation}
\begin{figure}[h]
\begin{minipage}{0.55\textwidth}
\psfig{figure=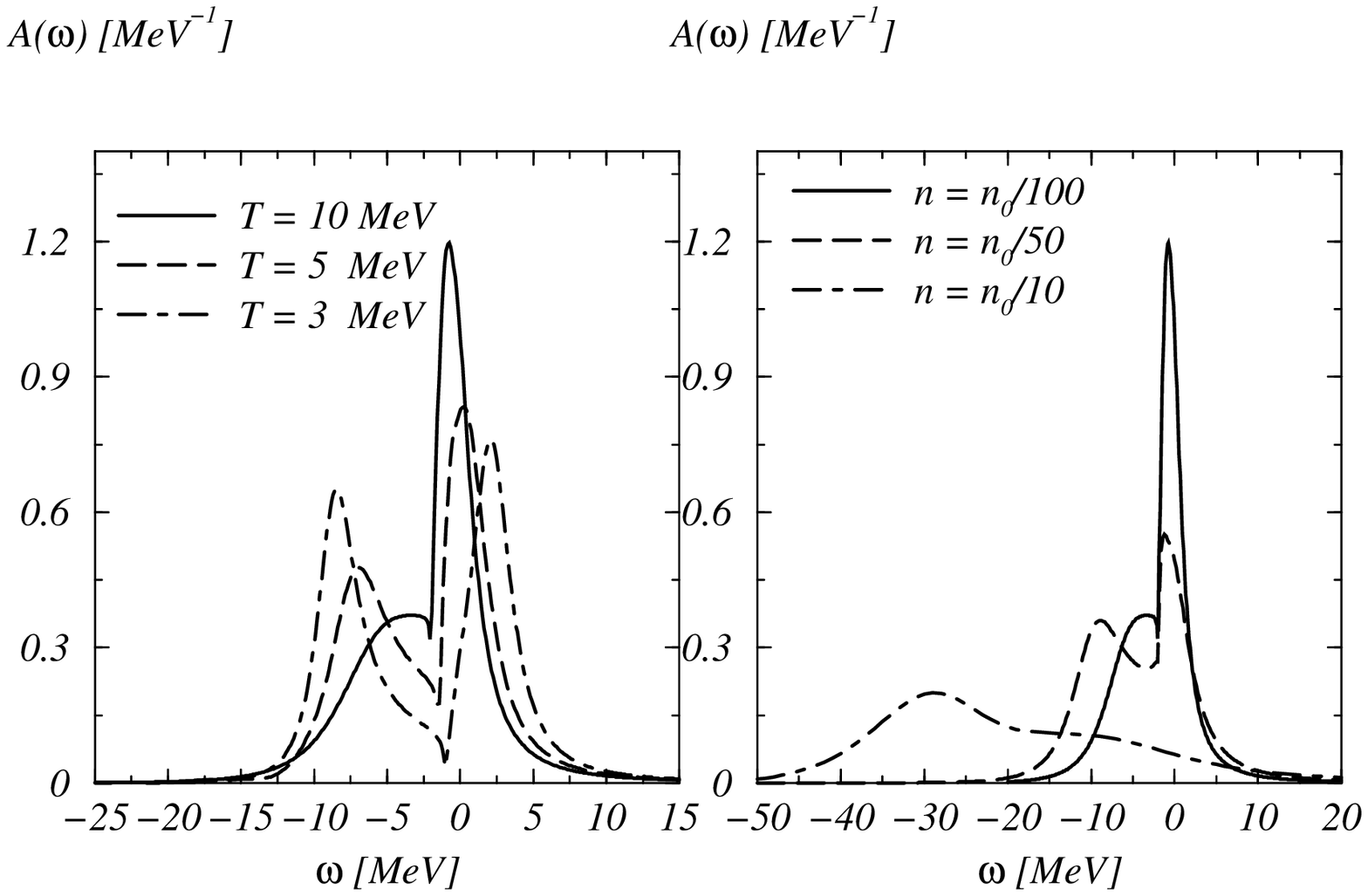,width=\textwidth}
\caption{\label{fig:spec} The spectral function $A_1(\omega)$
  for different temperatures and densities including two particle
  correlations. The deuteron bound state is recognized as the left
  shoulder. }
\end{minipage}
\hfill
\begin{minipage}{0.43\textwidth}
\psfig{figure=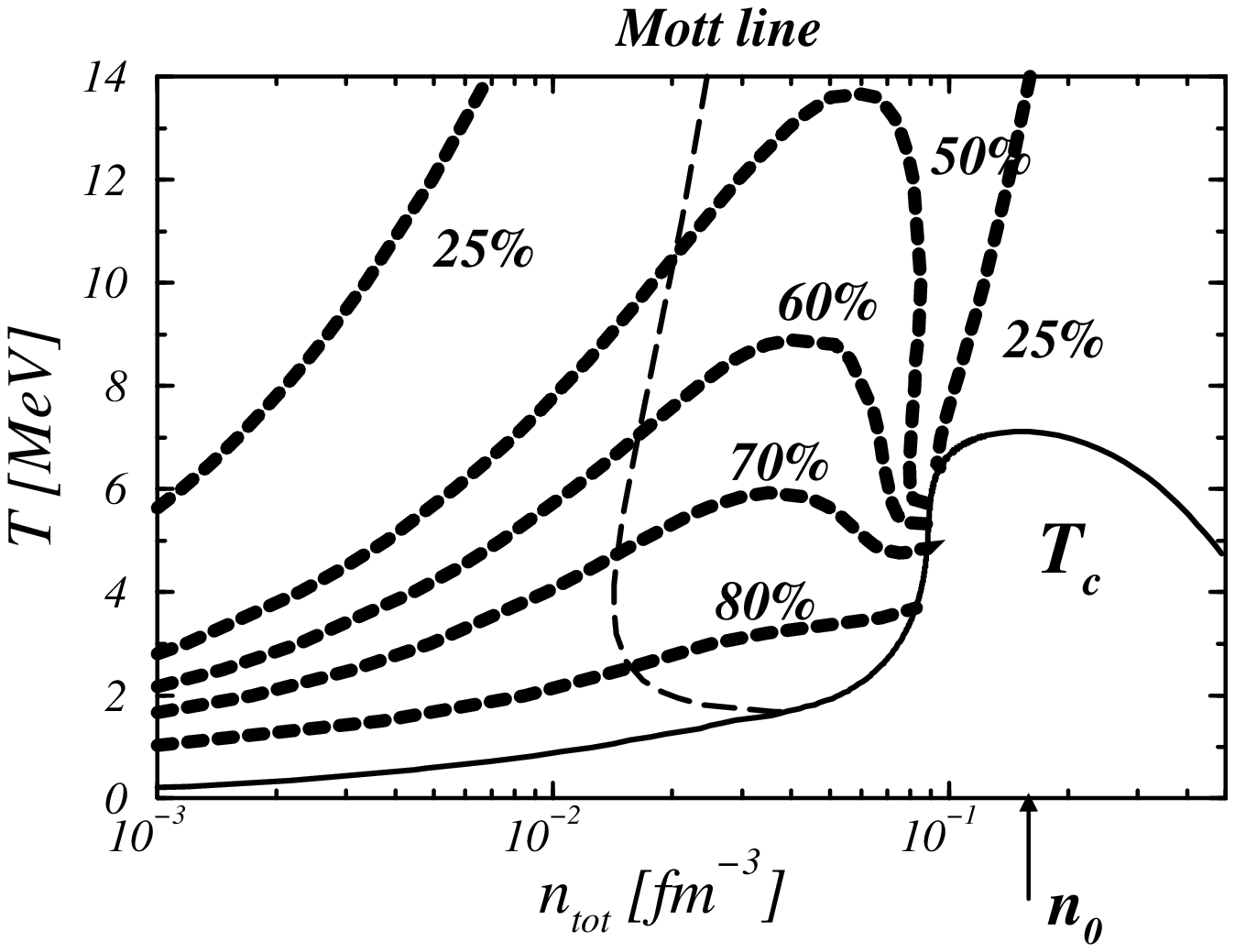,width=\textwidth}
\caption{\label{fig:corr} Temperature density plane for nuclear matter 
  as a function of the total density. The amount of correlates density
  is given in per cent of the total density.}
\end{minipage}
\end{figure}

As an example Figure~\ref{fig:spec} shows the spectral function using
the full two body t-matrix to describe the
correlations~\cite{schnell}.  As a consequence nuclear matter cannot
be considered as a system of independent quasi particles but for a
large part it is correlated up to full pairing in the superfluid
phase. This is depicted in Fig.~\ref{fig:corr}, where the dashed lines
show equal contribution of the correlated density to the total
density~\cite{stein}. The basis to treat correlated densities is
provided by a generalization of the Beth-Uhlenbeck
approach~\cite{sch90}. The nuclear density $n=n(\mu,T)$ is given by
\begin{equation}
n=n_{\rm free} + n_{\rm corr},\qquad n_{\rm corr} = 2n_2 + 3n_3 +
\dots
\end{equation}
where $n_{2,3}$ denotes the two, three-particle correlations, present
as bound/scattering states. In first iteration these correlations may
be treated on the basis of residual interactions between the quasi
particles. The exact two particle equations to be solved are known as
Bethe-Goldstone or Feynman-Galitski equations depending on some
details. In ladder approximation the equation for the two body Green
function reads
\begin{equation}
G_2(z) =   \frac{\bar f_1\bar f_2 - f_1f_2}
{z - \varepsilon_1 - \varepsilon_2}
+    \frac{\bar f_1\bar f_2 - f_1f_2}
{z - \varepsilon_1 - \varepsilon_2} \,V_2\, G_2(z),
\label{eqn:green2}
\end{equation}
where $\bar f=1-f$. Introducing the two-body $t$-matrix in a standard
fashion both bound and scattering states have been
solved~\cite{sch90}. 
\begin{figure}[tbh]
\begin{minipage}{0.48\textwidth}
\psfig{figure=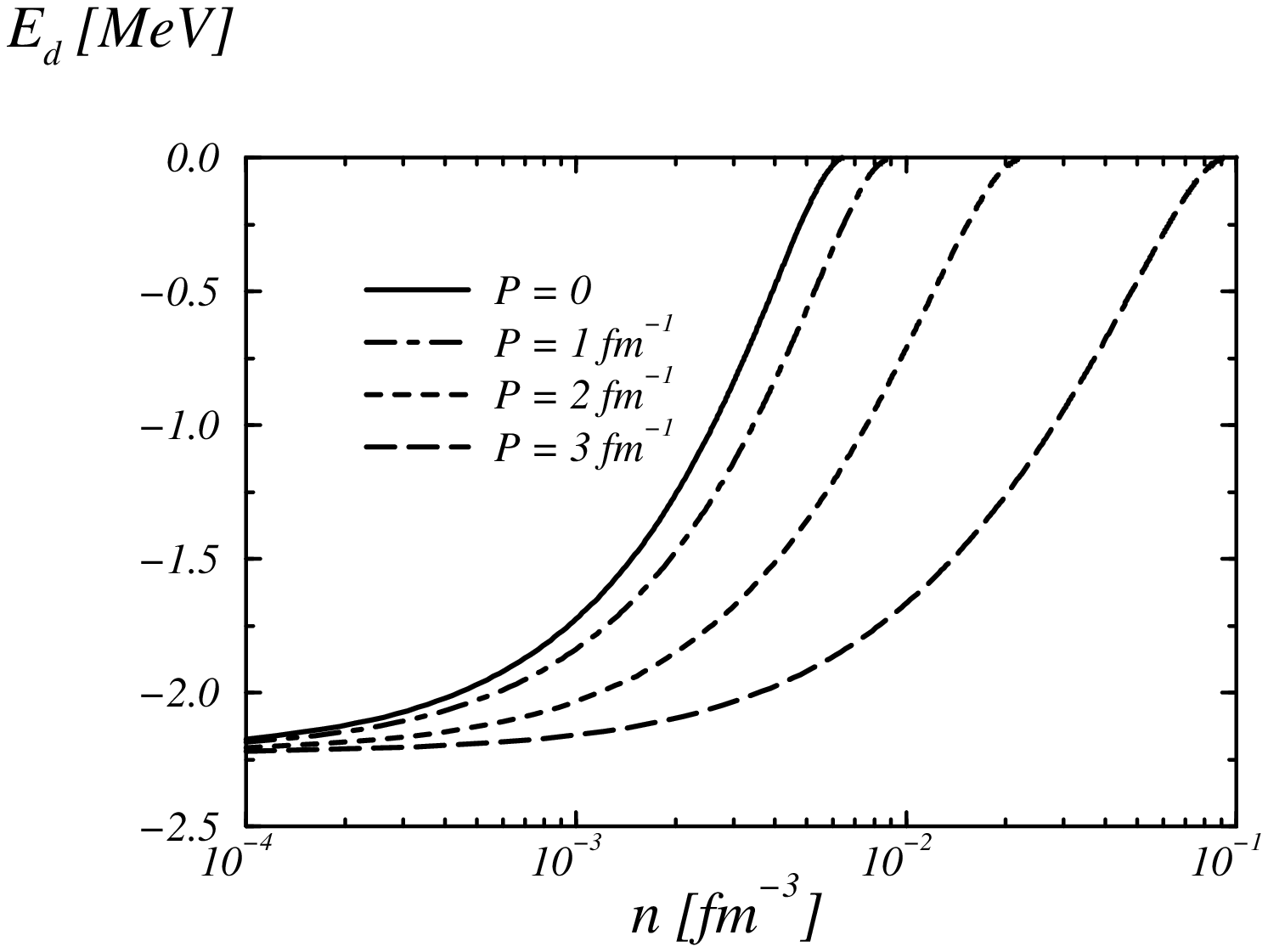,width=\textwidth}
\caption{\label{fig:deut} The deuteron binding energy as a function of 
  the nuclear density for $T=10$ MeV. $P$ denotes the relative momentum 
  between the deuteron and the medium.}
\end{minipage}
\hfill
\begin{minipage}{0.48\textwidth}
\psfig{figure=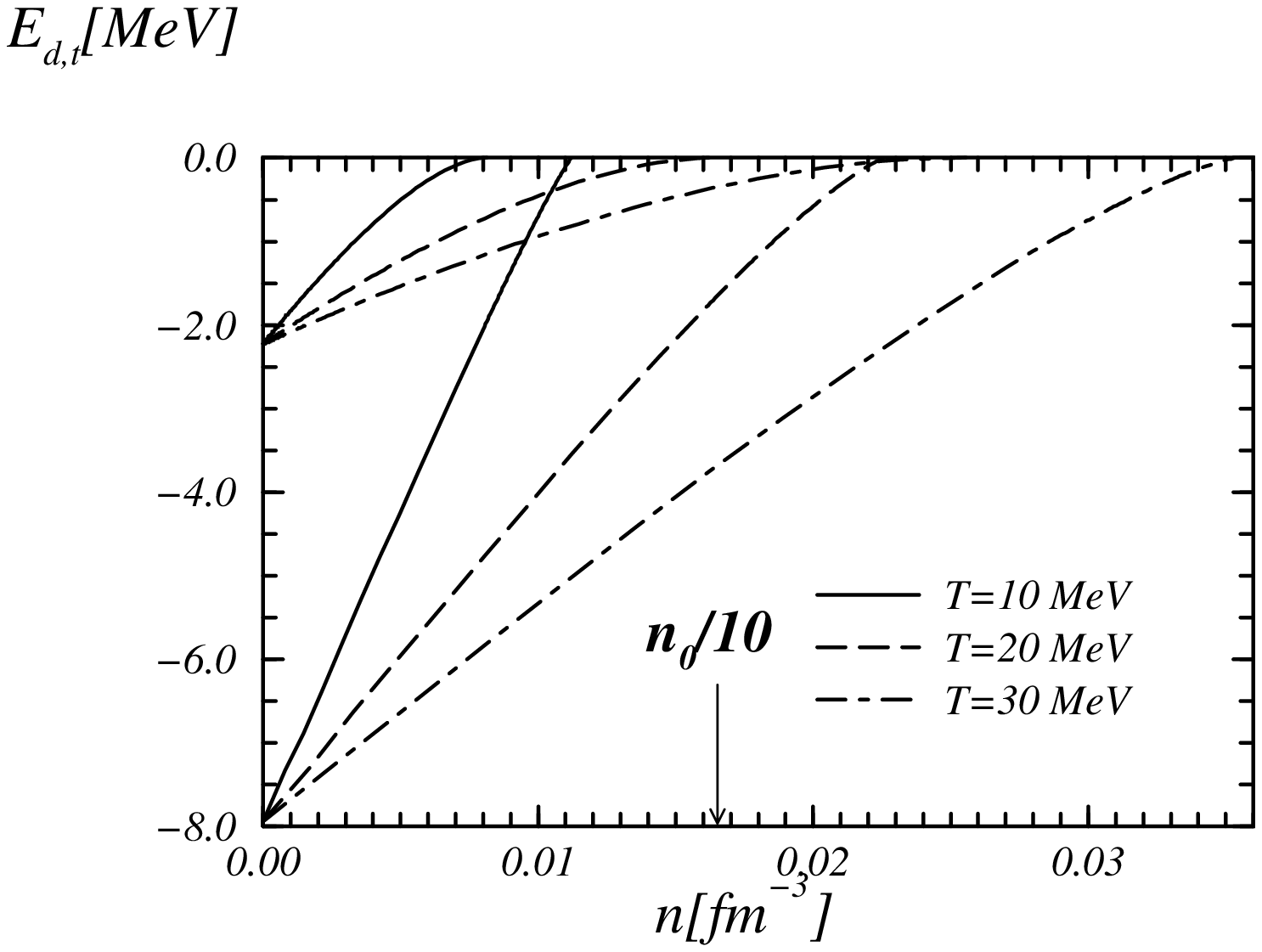,width=\textwidth}
\caption{\label{fig:triton} The triton binding energy as a function
  of nuclear density for different temperatures $T$. The triton rests
  in the medium. The respective $Nd$ continua are also indicated.}
\end{minipage}
\end{figure}

As an example the deuteron energy is shown in Fig.~\ref{fig:deut}. The
Mott density is defined through the condition $E_d=0$. Note that the
Fermi functions $f_1$ for particle 1 etc. depends on the relative
momentum $P$ between the deuteron and the medium. The respective
three-particle Faddeev type equation for the Green function has been
given in Ref.~\cite{bey96},
\begin{equation}
G_3(z)= \frac{\bar f_1\bar f_2\bar f_3 + f_1f_2f_3}
{z - \varepsilon_1 - \varepsilon_2- \varepsilon_3}
+\frac{(\bar f_1\bar f_2 - f_1f_2) V_2(12) + {\rm perm.}}
{z - \varepsilon_1 - \varepsilon_2- \varepsilon_3}
\;G_3(z)
\label{eqn:green3}
\end{equation}
Eq.~\ref{eqn:green3} has been rewritten using the AGS
approach~\cite{AGS}. The resulting in-medium AGS equations
are
\begin{equation}
U_{\ga\gb} = (1-\gd_{\ga\gb}) \left( 
\frac{\bar f_1\bar f_2\bar f_3 + f_1f_2f_3}
{z - \varepsilon_1 - \varepsilon_2- \varepsilon_3}\right)^{-1}+ 
\sum_{\gc\neq \ga}   T_3^{(\gc)} \frac{\bar f_1\bar f_2\bar f_3 + f_1f_2f_3}
{z - \varepsilon_1 - \varepsilon_2- \varepsilon_3} U_{\gc\gb}.
\label{eqn:AGS}
\end{equation}
where $T_3^{(\gc)}$ is the solution of the in-medium two-body problem, 
e.g. for $\gc=3$
\begin{equation}
T_3^{(3)}= (1-f_3+g(\varepsilon_1+\varepsilon_2))^{-1}
V_2+V_2\frac{\bar f_1\bar f_2 - f_1f_2}
{z - \varepsilon_1 - \varepsilon_2- \varepsilon_3}T_3^{(3)}.
\end{equation}
and the Bose function is given by $ g(\omega) = 1/(e^{\gb(\omega -
  2\mu)} - 1)$.  This equation has been solved for the $Nd$ reaction
relevant for deuteron formation~\cite{bey96,bey97} (see below),
assuming $f(\varepsilon)^2\ll f(\varepsilon)$, compare
Fig.~\ref{fig:fermi}.  Recently, the triton bound state equation has
also been solved~\cite{bey98}. The triton binding energy is shown in
Fig.~\ref{fig:triton}. The deuteron as well as the triton binding
energies weaken if the nuclear density is increased until the Mott
density is reached. This tendency is dominated by the Pauli blocking
of the surrounding medium.

The four-nucleon correlation is believed to play a significant role
for lower densities and temperatures. Exploratory calculations using a
simple variational ansatz for the $^4$He wave function predict an
$\alpha$ condensate/quartetting on top of the deuteron
condensate/triplet pairing that leads to superfluidity~\cite{roe98}.

Bose systems behave quite differently with respect to the occurrence
of bound states.  The bose functions enhance the effective residual
interaction that might lead to an ``opposite Mott effect'', i.e.
existence of bound states and also pairing (condensate) even if no
bound state exist for the isolated case. An example is provided by a
pion gas, were a pionic condensate may occur~\cite{alm97} that are
discussed, e.g. in the context of neutron stars~\cite{pin92}.

\section{Reactions}
Nuclear reaction rates play an essential role in the formation of
stars like the sun. The standard solar model is based on binary
collisions. Recently, triple reactions, e.g. $e+ ^3{\rm He} + \alpha
\rightarrow ^7 {\rm Be} + e$ to be compared to $^3{\rm He} + \alpha
\rightarrow ^7 {\rm Be} + \gamma$ have been investigated and found to
be rather small in plasmas at solar conditions~\cite{belayev}.
However, note that triple collisions are mostly non-radiative and that
they may be more important for other stars than the sun or at the
early universe~\cite{belayev}.

Another example are dense ionic plasmas, where the ionisation rate
depends on three-particle reactions that are presently treated in Born
approximation. Since the residual interaction is Coulombic this may be
considered a good approximation and it reproduces the experimental
results for hydrogen like plasmas~\cite{bornath}. For higher ionized
plasmas this might not be the case and the application of Faddeev like
methods may be in order. These will be sketched in the following for
nuclear matter.

\begin{figure}[tbh]
\begin{minipage}{0.48\textwidth}
\psfig{figure=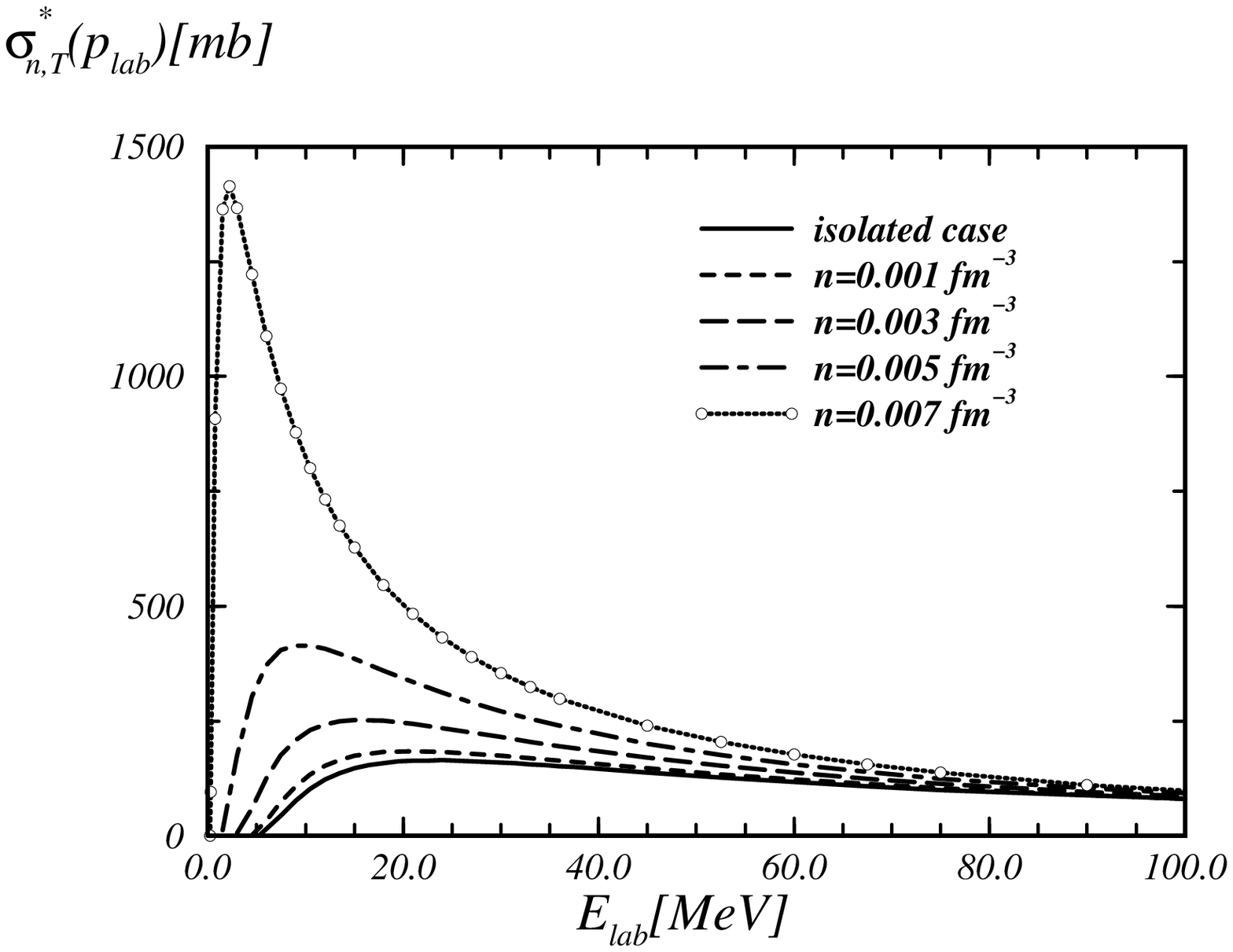,width=\textwidth}
\caption{\label{fig:cross} In-medium break-up cross section at $T=10$
  MeV. Isolated cross section is shown as solid line. Other lines are 
  due to different nuclear densities.  }
\end{minipage}
\hfill
\begin{minipage}{0.48\textwidth}
\psfig{figure=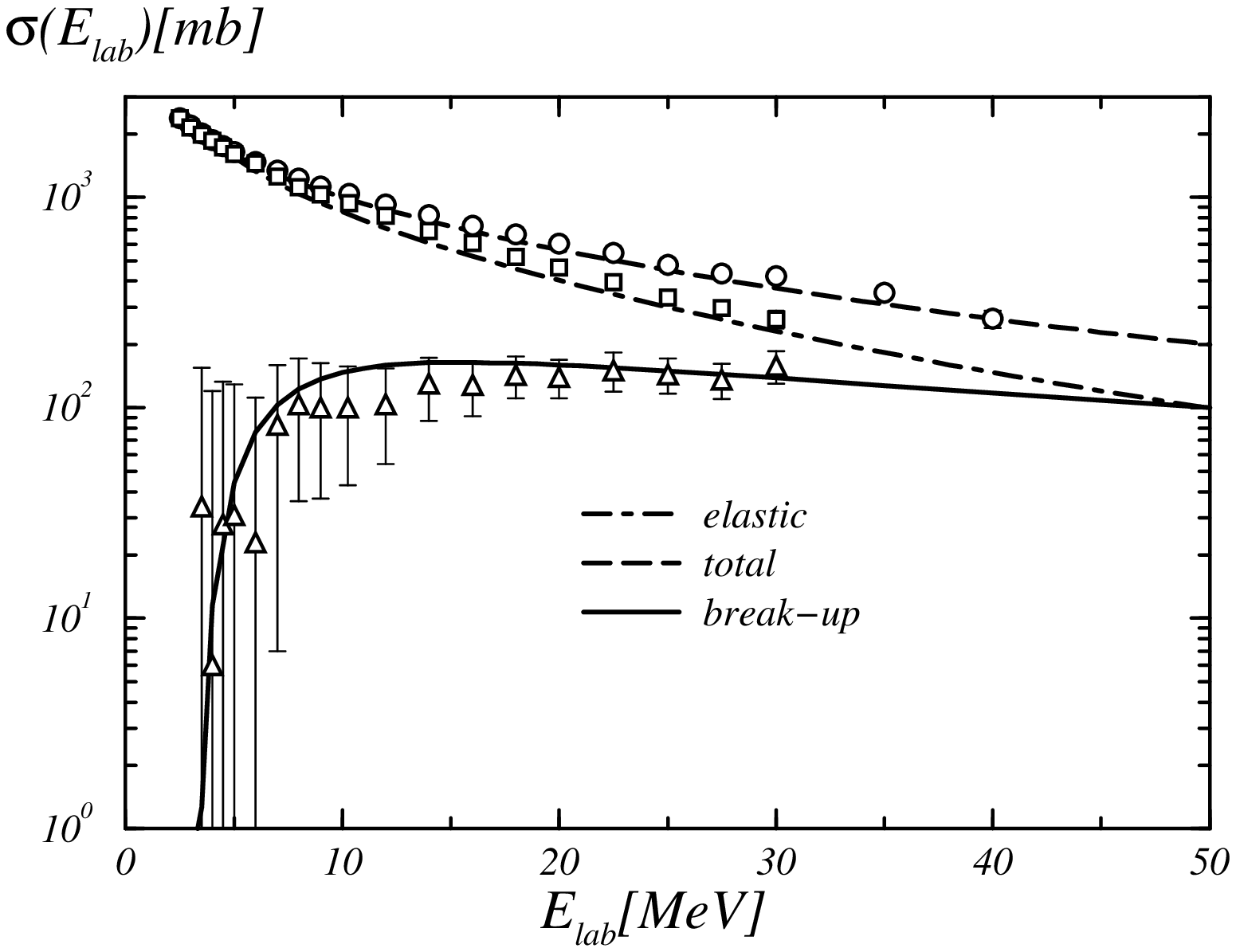,width=\textwidth}
\caption{\label{fig:isocross}  A comparison of the total, elastic,  
  and break-up cross sections $nd\rightarrow nd$, $nd\rightarrow
  nnp$ with the experimental data of
  Ref.~\protect{\cite{sch83}}.}
\end{minipage}
\end{figure}

The generalized quantum kinetic Boltzmann equations for the
nucleon $f_N(p,t)$ (momentum $p$) and deuteron $f_d(P,t)$ (momentum $P$)
distribution functions~\cite{zub96}
\begin{eqnarray}
f_N(p,t)&=&\langle a^\dagger_{Np} a_{Np}\rangle
\equiv {\rm Tr}\{\rho(t)  a^\dagger_{Np} a_{Np}\}\nonumber\\
f_d(P,t)&=&\langle b^\dagger_{dP} b_{dP}\rangle
\end{eqnarray}
are coupled and read
\begin{eqnarray}
f_N(p,t)&=&-{\cal D}_N(p,t) + {\cal I}_N(p,t)\nonumber\\
f_d(P,t)&=&-{\cal D}_d(P,t) + {\cal I}_d(P,t),
\end{eqnarray}
The first term reflects the so called Vlasov term and is related to
the mean field. The second term is the collision term that is
responsible for equilibration of the system. The explicit form of the
integral ${\cal I}_N(p,t)$ is
\begin{equation}
{\cal I}_N(p,t)= {\cal I}^>_N(p,t)f_N(p,t) 
- {\cal I}^<_N(p,t)\bar f_N(p,t),
\end{equation}
 where, e.g. ${\cal I}^>_N(p,t)$ is given by
\begin{eqnarray}
{\cal I}^>_N(p,t)&=&\int dk dk_1dk_2 
\left|\langle kp|T_{NN\rightarrow NN}|k_1k_2\rangle\right|^2
\bar f_N(k_1,t)\bar f_N(k_2,t) f_N(k,t) \nonumber\\
&&+\int dk dk_1dk_2 dk_3
\left|\langle kp|U_{Nd\rightarrow NNN}|k_1k_2k_3\rangle\right|^2
\nonumber\\
&&\qquad\qquad
\times\bar f_N(k_1,t)\bar f_N(k_2,t)\bar f_N(k_3,t) f_d(k,t)+\dots
\end{eqnarray}
The solution of this equation are the distribution functions $f_N$ and
$f_d$, that however also appear in the transition matrices
$T_{NN\rightarrow NN}$, $U_{Nd\rightarrow NNN}$, etc. Therefore a full
solution of this equation is a difficult problem. For small
fluctuations from the equilibrium distributions the equations may be
linearized in the framework of linear response theory. The binary
collision approximation may the transition matrix elements depend on
the equilibrium distribution only and the results of the previous
sections can be utilized. Even this in-medium dependence has hardly
been considered in modeling of heavy ion collision~\cite{alm95A}.
Also, for three-nucleon collisions so far only the impulse
approximation has been used~\cite{dan91}. Here we solve the in-medium
Faddeev type equation that includes the Hartree-Fock self energy shift
and the Pauli blocking in a consistent way, Eq.~\ref{eqn:green3}. The
resulting break-up cross section for a typical temperature of $T=10$
MeV and densities below the deuteron Mott density is shown in
Fig.~\ref{fig:cross}. For a comparison of the quality of the model the
isolated cross section along with the experimental data~\cite{sch83}
are shown in Fig.~\ref{fig:isocross}.
From inspection of Fig.~\ref{fig:cross} we see that the in-medium cross 
section is significantly enhanced compared to the isolated on. The
threshold is shifted to smaller energies, which is because the binding 
energy of the deuteron becomes smaller. We observe that for higher
energies the medium dependence of the cross section becomes much
weaker. 

\begin{figure}[bh]
\begin{minipage}{0.58\textwidth}
\psfig{figure=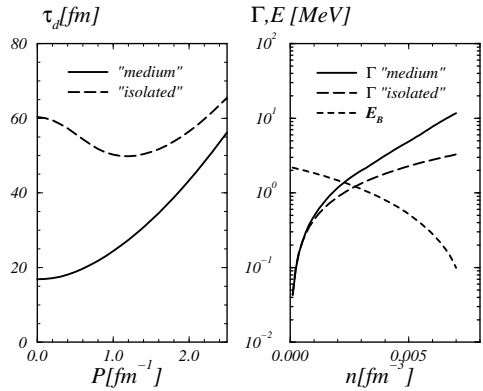,width=0.9\textwidth}
\end{minipage}
\hfill
\begin{minipage}{0.4\textwidth}
\caption{\label{fig:life} Fluctuation time of the deuteron
  distribution as a function of the deuteron momentum $P$. As input
  the in-medium cross section is compared to the isolated one, nuclear 
  density $n=0.007$ fm$^{-3}$, temperature $T=10$ MeV.}
\end{minipage}\vfill
\end{figure}
Though the change of the cross section looks dramatic, the
quantum Boltzmann equation still has some additional medium dependence 
that may change this effect in the observables $f_N$ and $f_d$. Within 
the linear response theory it is possible to calculate the chemical
reaction time due to the break-up process. For small fluctuations
$\gd f(t)$ linear response leads to
\begin{equation}
\partial_t \gd f_d^{\rm reaction}(P,t) = \frac{1}{\tau_P}
\gd f_d^{\rm reaction}(P,t)
\end{equation}
where the ``life time'' of deuteron fluctuations has been
introduced,
\begin{equation}
\tau^{-1}_d 
= \frac{4}{3!} \int dk_Ndk_1 dk_2 dk_3\;
\left|\langle kp|U_{Nd\rightarrow NNN}|k_1k_2k_3\rangle\right|^2
\; \bar f_{1}\bar f_{2}\bar f_{3}f_{\varepsilon}
\;2\pi\gd(E - E_0).
\label{eqn:lifetime}
\end{equation}
which can be related to the break-up cross section given in
Eq.~\ref{fig:cross}. For low densities the life time (as a function of
the deuteron momentum $P$) and the inverse life time, i.e. the width,
at $P=0$ along with the deuteron binding energy for comparison is
shown in Fig.~\ref{fig:life}~\cite{bey97}. These times have to be
compared to the approximate duration of the heavy ion collision of
about 200 fm.

\section{Conclusion}
The treatment of correlations and triple collisions in non-ideal many
particle quantum systems opens up a new field for few-body methods.
The examples shown have been mostly from nuclear physics.  However,
applications are possible for stellar plasmas to improve the
description of the basic quantity, which is the spectral function and
to include correlations into the equation of state. In the laboratory
the ionisation rate of dense ionic plasmas is determined by
three-particle collisions. A description is presently restricted to
hydrogen like plasmas, where the Born approximation for the
three-particle reaction is sufficient. Typical applications in nuclear
physics are related to heavy ion collisions, here in particular the
formation of light clusters such as deuterons, helium-3, tritium, and
alpha particles. The conditions are satisfied during the final stage
of the heavy ion collision, where a temperature of $T\simeq 5\dots 10$
MeV may be meaningful and the densities are below the Mott densities
of cluster formation. The results are also relevant for the equation
of states of neutron stars.

The approach given here follows the quantum statistical description as
it provides a rigorous, systematic treatment of many particle systems.
The major approximation utilized is the cluster expansion to decouple
the infinite hierarchy of equations (Green functions or kinetic
equations). This approximation clearly goes beyond the quasi particle
picture as it includes the residual interactions in a systematic way.
This is done rigorously using few-body methods. To this end few-body
equations have to be substantially generalized. Presently, these
equations resemble an RPA structure~\cite{sch73}, however extended to
finite temperatures.  

The validity of this already ambitious approach has to be checked by
facing experimental results. This seems easier for ionic plasmas, e.g.
to calculate the ionisation rate, or for electron-hole plasmas in the
context of exciton formation.  Testing the validity of the approach
for nuclear physics needs a handle of heavy ion collisions.  Presently
one relays on numerical simulations of the complicated dynamics of a
heavy ion collision, which is subject to discussions by its own.
Typical heavy ion simulation codes that require microscopic input are
based on e.g. a Boltzmann-Uehling-Uhlenbeck treatment or on quantum
molecular dynamics. The results presented here may however also be
relevant for standard nuclear physics, e.g. electron scattering off
heavy nuclei, when correlations are considered, and one therefore
needs to go beyond the quasi particle picture.

\begin{acknowledge}
  It is a pleasure to thank my colleagues in Rostock who provided me
  with some material presented during the talk.
\end{acknowledge}


\SaveFinalPage
\end{document}